\documentclass[aps,twocolumn,nofootinbib,preprintnumbers,groupedaddress,letterpaper]{revtex4}

\usepackage[dvips]{color}
\usepackage[normalem]{ulem}
\usepackage{amsmath}
\usepackage{enumerate}
\usepackage{amsfonts}
\usepackage{epsfig}
\usepackage{yfonts}

\usepackage{graphicx}
\usepackage{graphicx}
\usepackage{bm}
\usepackage{subfigure}

\newcommand\comment[1]{}

\newcommand\poincare{Poincar\' e }

\def\le{\left}
\def\ri{\right}

\def\({\left(}
\def\){\right)}
\def\[{\left[}
\def\]{\right]}
\def\<{\langle}
\def\>{\rangle}

\newcommand\half{{\ensuremath{\frac{1}{2}}}}
\newcommand\p{\ensuremath{\partial}}

\newcommand\field[1]{{\ensuremath{\mathbb{{#1}}}}}

\newcommand{\RR}{\field{R}}

\newcommand{\be}{\begin{equation}}
\newcommand{\ee}{\end{equation}}
\newcommand{\bea}{\begin{eqnarray}}
\newcommand{\eea}{\end{eqnarray}}
\newcommand{\bwt}{\begin{widetext}}
\newcommand{\ewt}{\end{widetext}}

\newcommand{\bi}{\begin{itemize}}
\newcommand{\ei}{\end{itemize}}
\newcommand{\ben}{\begin{enumerate}}
\newcommand{\een}{\end{enumerate}}
\newcommand{\bca}{\begin{cases}}
\newcommand{\eca}{\end{cases}}
\newcommand{\bln}{\begin{align}}
\newcommand{\eln}{\end{align}}
\newcommand{\bst}{\begin{split}}
\newcommand{\est}{\end{split}}

\begin{document}

\preprint{CERN-PH-TH-2015-209}

\title{The broken string in anti-de Sitter space}

\author{David Vegh}
\email{dvegh@ias.edu}

\affiliation{\it  Institute for Advanced Study, Princeton, NJ 08540, USA \\ }
\affiliation{\it Theory Group, Physics Department, CERN, CH-1211 Geneva 23, Switzerland}

\date{\today}

\begin{abstract}

This paper describes an efficient method for solving the classical string equations of motion in (2+1)-dimensional anti-de Sitter spacetime. Exact string solutions are identified that are the analogs of piecewise linear strings in flat space. They can be used to approximate any smooth string motion to arbitrary accuracy. Cusps on the string move with the speed of light and their collisions are described by a Picard-Lefschetz-type formula.
Explicit examples are shown with the string ending on two boundary quarks.
The technique is ideally suited for numerical simulations. A Mathematica notebook that has been used to generate the relevant figures is also included.

\end{abstract}

\maketitle

\section{Introduction}

This paper is concerned with the classical motion of a long string in (2+1)-dimensional anti-de Sitter spacetime. If the string is open, it stretches between two points on the boundary. According to the AdS/CFT correspondence \cite{Maldacena:1997re, Gubser:1998bc, Witten:1998qj}, the dual boundary gauge theory contains a Wilson loop on which the string ends \cite{Maldacena:1998im, Rey:1998ik}. It is useful to think of the Wilson loop as the path of an infinitely heavy quark-antiquark pair. The string in the bulk is the holographic dual to the color flux tube that connects them.
The motion of the quarks can be specified. Generic motion creates non-linear waves on the string that propagate toward the other endpoint. The aim of this paper is to describe the collisions of these waves and therefore calculate the string motion in the most efficient way.

The canonical embedding of AdS$_3$ into $\RR^{2,2}$ is given by the universal covering space of the surface
\be
  \label{eq:surface}
  \vec Y \cdot \vec Y \equiv -Y_{-1}^2 - Y_0^2 + Y_1^2 + Y_2^2 = -1 .
\ee
Time corresponds to the phase on the $Y_{-1}$, $Y_0$ plane. Since on the surface this time dimension is compact, the surface itself only covers a part of global AdS.

Coordinates on the \poincare patch will also be used, for which the metric is
\be
  \nonumber
  ds^2 = {-dt^2 + dx^2 + dz^2 \over z^2}
\ee
The coordinate transformation is given by
\be
  \nonumber
  (t, \, x, \, z) = \le(  {Y_{0} \over Y_{2} - Y_{-1}}, \ {Y_{1} \over Y_{2} - Y_{-1}}, \ {1 \over Y_{2} - Y_{-1}} \ri).
\ee

The string motion is described by a classically integrable system.
The string equations of motion in conformal gauge are
\be
  \label{eq:eoms}
  \p \bar\p \vec Y - (\p \vec Y \cdot \bar\p \vec Y ) \vec Y = 0 .
\ee
These are supplemented by the Virasoro constraints
\be
  \nonumber
  \p \vec Y \cdot \p \vec Y = \bar\p \vec Y \cdot \bar\p \vec Y = 0 .
\ee
The above system can be reduced to a generalized  sinh-Gordon theory \cite{Pohlmeyer:1975nb, DeVega:1992xc, Jevicki:2007aa, Jevicki:2009uz, Alday:2009yn, Irrgang:2012xb} by defining
\be
  \nonumber
  e^{2\alpha(z,\bar z)} = \half \p \vec Y \cdot \bar\p \vec Y
\ee
\be
  \nonumber
  N_a = \half e^{-2\alpha} \epsilon_{abcd} Y^b \p Y^c \bar\p Y^d
\ee
\be
  \nonumber
  p = -\half \vec N \cdot \p^2 \vec Y \ , \qquad  \bar p = \half \vec N \cdot \bar\p^2 \vec Y \ .
\ee
Note that $\vec N \cdot \vec Y =\vec N \cdot \p \vec Y = \vec N \cdot \bar\p \vec Y  = 0 $ and $\vec N \cdot \vec N = 1$. Furthermore, $p=p(z)$ and $\bar p = \bar p(\bar z)$ are holomorphic and antiholomorphic functions, respectively. Then, the potential $\alpha$ satisfies the generalized sinh-Gordon equations
\be
  \label{eq:sinh}
  \p \bar\p \alpha(z,\bar z) - e^{2\alpha} + p(z)\bar p(\bar z) e^{-2\alpha} = 0 .
\ee
Given a solution, the string embedding can be computed by solving an auxiliary fermion scattering problem where $\alpha$ appears as a potential. This is feasible\footnote{For an analytical solution to the scattering problem with N singular  solitons, see \cite{Jevicki:2009uz}.} but not very practical for numerical calculations.

In this paper, we follow a different approach. Highly symmetric string pieces will be glued together; these are the AdS analogs of straight lines. This way any generic solution can be approximated. The corresponding potential satisfies (\ref{eq:sinh}) with $p(z)=0$, i.e.  the Liouville equation. At the attachment points the string will have cusps.
Information about the original sinh-Gordon equation is condensed to these points.

In the next section, the basic symmetric solution is described. In section III, the gluing procedure is explained. Section IV discusses what happens when two cusps collide. Section V contains various examples. Details about the numerical calculations are presented in the appendix.

\begin{figure}[h]
\begin{center}
\includegraphics[scale=0.45]{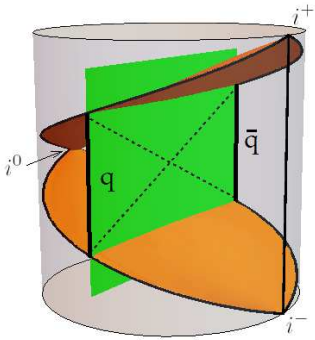}
\caption{\label{fig:globalads}
The basic string solution (in green) in global AdS$_3$ spacetime. The string ends on two quarks ($q$ and $\bar q$) on the boundary. The \poincare patch is bounded by the two orange surfaces.
}
\end{center}
\end{figure}

\section{The basic solution}

\begin{figure}[h]
\begin{center}
\includegraphics[scale=0.53]{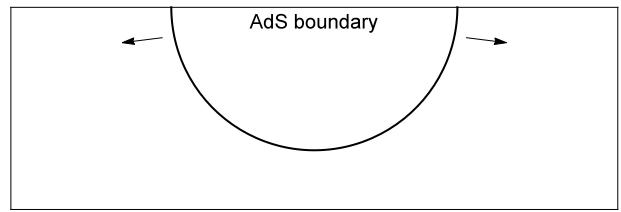}
\caption{\label{fig:basicstring} The basic string solution (thick line) on the \poincare patch. The horizontal and vertical axes are the $x$ and $z$ coordinates, respectively. The top of the figure is $z=0$, the AdS boundary.
The quark and the antiquark move on this line in opposite directions and are connected by a circular string.
}
\end{center}
\end{figure}

\begin{figure}[h]
\begin{center}
\includegraphics[height=4.5cm]{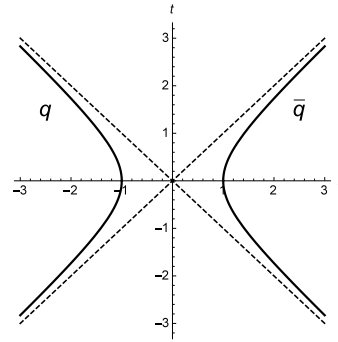}
\caption{\label{fig:motion} Motion of the string endpoints in the boundary Minkowski spacetime. The acceleration of the quarks is constant and they are out of causal contact as indicated by the light rays (dashed lines).
}
\end{center}
\end{figure}

A simple solution to the string equation of motion is obtained by setting $N$ to be a constant unit length vector. This is the AdS$_3$ analog of a static infinite straight string in flat spacetime.
By applying an appropriate transformation from the $SO(2,2)$ isometry group of  AdS$_3$, $\vec N$ can be rotated such that
\be
  \nonumber
  \vec N(t) = (0,0,0,1)
\ee
The corresponding spacetime solution is an infinitely long string: AdS$_2$ embedded into AdS$_3$. FIG. \ref{fig:globalads} shows the worldsheet in global AdS spacetime. The string in these coordinates is static and ends on two antipodal points on the boundary $S^1$.
FIG. \ref{fig:basicstring} shows the string on the \poincare patch at a given  time $t$.
The string is a contracting/expanding semi-circle centered at $(x,z)=(0,0)$ with radius $R(t) = \sqrt{1 + t^2}$.
On the boundary of AdS, the two quarks move on hyperbolae: $x_{1,2}(t) = \pm R(t)$, see FIG. \ref{fig:motion}.

$\vec N$ will be called the normal vector, or the vector perpendicular to the string patch.
Since $\vec N^2 = 1$, the AdS$_2$ spaces form a 3-parameter set. The string solutions are still shrinking/expanding semicircles, but they are now shifted in the $x$ and $t$ directions, and their minimal radii are also different.
Let us denote the center of the circle by $x_0$, the minimal radius by $R_0$, and the time when the string reaches the minimal radius by $t_0$.
The relationship between $\vec N$ and these parameters is given by
\be
  \nonumber
  (t_0, \, x_0, \, R_0) = \le(  {-N_{0} \over N_{-1} + N_2}, \ {-N_{1} \over N_{-1} + N_2}, \ {1 \over | N_{-1} + N_2 |} \ri) .
\ee

Let us briefly discuss how this solution is connected to the $\alpha=0$ ``ground state'' of the sinh-Gordon theory. This trivial solution corresponds to a rotating string \cite{Gubser:2002tv} in the infinite angular momentum limit. The endpoints are on the boundary of AdS and they move with the speed of light. In order to stop the rotation, two anti-solitons can be added to the potential. In the center-of-mass frame \cite{Jevicki:2007aa},
\be
  \label{eq:2sols}
  \alpha(z,\bar z) = -\half \log\le( {v_0 \cosh X - \cosh T \over v_0 \cosh X + \cosh T } \ri)^2
\ee
where $X = 2{z+\bar z \over \sqrt{1-v_0^2}}$, $T = 2{v_0(z-\bar z )\over \sqrt{1-v_0^2}}$, and $v_0$ is the initial speed of the anti-solitons. $\alpha$ satisfies (\ref{eq:sinh}) with $p=\bar p = 1$.

The metric conformal factor $e^{2\alpha} \to \infty$ at the location of the anti-solitons. Thus, the string touches the boundary at these points. On the worldsheet, the anti-solitons approach each other and after reaching a $d$ minimal  coordinate distance, they turn around.

In the $v_0\to 1$ limit,  $d~\propto~1-v_0$. By a simultaneous rescaling of the worldsheet coordinates $z \to \le({1-v_0 \over 2} \ri)z'$, we can zoom in on the collision point. Then, the anti-solitons will move on the hyperbola defined by $z'\bar z' = 1$ and the potential becomes
\be
  \nonumber
  \alpha(z' ,\bar z') = -\half\log \, (1-z' \bar z')^2 -\log \le( {1-v_0 \over 2} \ri) + \mathcal{O}(1-v_0)
\ee
The first term is a renormalized potential that satisfies  (\ref{eq:sinh}) with $p=0$.
The corresponding string embedding is the circular string with a constant normal vector.

\begin{figure}[h]
\begin{center}
\includegraphics[scale=0.53]{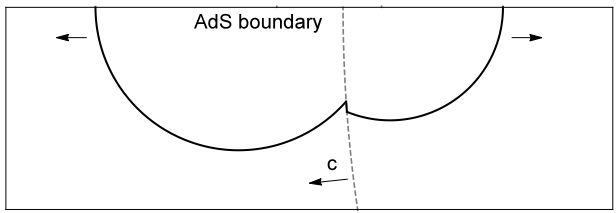}
\caption{\label{fig:lightspeed} String embedding if one of the quark velocities is suddenly changed. The string consists of three circular pieces. The middle one is badly behaved: it propagates with the speed of light (its center is on the boundary point where the quark was kicked). No information can pass through this piece in finite time.
}
\end{center}
\end{figure}

\section{Cusps on the string}

In order to describe more general string configurations, the circular string solution from the previous section must be perturbed. Waves moving in a single direction may easily be obtained following \cite{Mikhailov:2003er}.
The motion of one of the string endpoints on the boundary is specified by giving a one-parameter family of lightlike vectors $\vec l(\tau) \in \RR^{2,2}$. The string solution is given by
\be
  \nonumber
  \vec Y(\tau, \sigma) = -\dot {\vec l} + \sigma \vec l(\tau)
\ee
The induced metric on the string is
\be
\nonumber
 g = \left(
\begin{array}{cc}
a(\tau)^2 - \sigma^2 & 1 \\
1 & 0
\end{array}
\right)
\ee
where $a$ is the acceleration of the boundary quark.
The Ricci scalar is constant on the worldsheet.

The quark motion must be continuous so that the string worldsheet is not ripped apart. If the quark velocity jumps, then the solution will contain a badly behaved piece: an arc of string on the \poincare patch that propagates with the speed of light (see FIG. \ref{fig:lightspeed} and also \cite{Chernicoff:2013iga}). No perturbation can propagate through this piece, because the speed of the perturbation parallel to the string will be zero. The solution may be interpreted as a string that ends here.

\begin{figure}[h]
\begin{center}
\includegraphics[height=5cm]{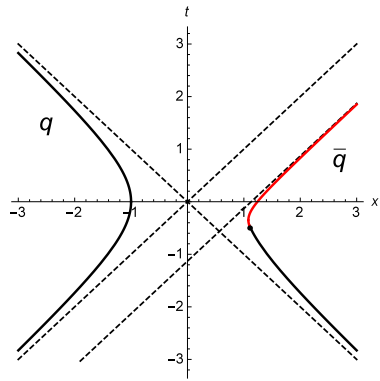}
\caption{\label{fig:motioncusp} Motion of the string endpoints in the boundary Minkowski spacetime  if one of the quark accelerations is suddenly changed. The quark worldline now asymptotes to a different light ray.
}
\end{center}
\end{figure}

If the acceleration of the quark changes continuously, then the string worldsheet will be smooth. Generic infalling (left-moving) waves are produced this way. However, collisions between left- and right-moving waves are not easily described. Therefore, in the following we will be interested in elementary building blocks whose shape survives the collisions without any deformations. These solitonic excitations are cusps on the string produced by letting the acceleration of the boundary quark jump in time\footnote{In principle, one could consider adding extra solitons on top of the double anti-soliton background of (\ref{eq:2sols}). However, the corresponding cusps on the string disappear in the $v\to 1$ limit.}. This is shown in FIG.  \ref{fig:motioncusp}.

\begin{figure}[h]
\begin{center}
\includegraphics[scale=0.53]{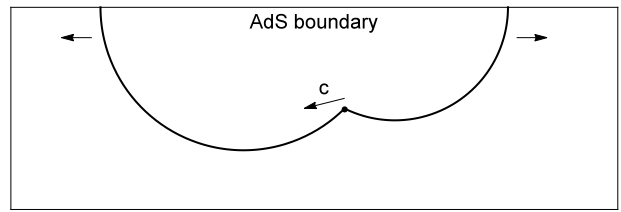}
\caption{\label{fig:1cusp} String embedding if one of the quark accelerations is suddenly changed. This creates a cusp that propagates on the string with the speed of light.  The string consists of two circular pieces. Their normal vectors are not arbitrary; they satisfy $\vec N_1 \cdot \vec N_2 = 1$.
}
\end{center}
\end{figure}

The resulting cusp moves on the string either left or right with the speed of light and connects two string patches, see FIG. \ref{fig:1cusp}.

If the worldline of the cusp is given by $\vec X(\lambda)$ and the normal vector of the left patch is denoted  $\vec N_L$, then the following equations are satisfied
\be
  \nonumber
   \vec N_L^2  = 1 \, , \qquad \vec X_\lambda \cdot \vec N_L = 0
\ee
\be
  \nonumber
  \vec X_\lambda^2 = -1 \, , \qquad   (\p_\lambda \vec X_\lambda)^2 = 0
\ee
The worldline parameter $\lambda$ can be chosen such that $\p_\lambda \vec X_\lambda$ is a constant vector. This is possible, because (\ref{eq:surface}) is a doubly ruled surface.
One can prove that on the $x-z$ plane (on the \poincare patch) the cusp moves on a straight line with the speed of light.
The vector perpendicular to the right patch can be expressed as
\be
  \nonumber
  \vec N_R = \vec N_L + \kappa \p_\lambda \vec X_\lambda
\ee
where $\kappa$ parametrizes the ``strength'' of the cusp.
If $\vec N_L$ is fixed, then the possible $\vec N_R$'s form a 2-parameter subset of the 3-parameter family of solutions which have a constant normal vector. In summary, they satisfy
\be
  \nonumber
   \vec N_L^2  = 1 \, , \qquad  \vec N_R^2  = 1 \, , \qquad  \vec N_L\cdot \vec N_R  = 1
\ee

\begin{figure}[h]
\begin{center}
\includegraphics[height=3.3cm]{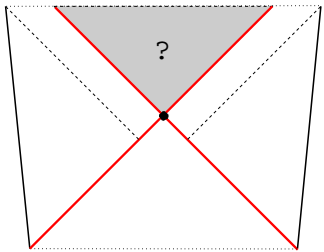}
\caption{\label{fig:penrosecoll} Worldsheet Penrose diagram corresponding to the collision of two waves (in red) that were sent in from the boundary in the far past. The string embedding in the gray region is to be computed.
}
\end{center}
\end{figure}

\section{Collision of cusps}

The cusps on the string propagate with the speed of light and (in global AdS) they inevitably collide.
The worldsheet Penrose diagram of FIG. \ref{fig:penrosecoll} illustrates the situation: the string patch between two cusps (red lines) disappears and after the collision a new patch is created (in gray).

The string embedding is shown in FIG. \ref{fig:expl} (thick line). Very close to the collision point, spacetime is approximately flat and the three string pieces are straight lines labeled $N_1$, $A$ and $N_2$ that move with constant perpendicular velocities. The corresponding normal vectors will be denoted by $\vec N_1$, $\vec A$ and $\vec N_2$. There exists an $SO(2,2)$ transformation (or an analogous boost in flat spacetime) that puts us in a frame where $N_1$ and $N_2$ are static and $A$ moves with the velocity $v$.  The configuration is highly symmetric and the only thing that can happen is that the collision switches $\vec v \to -\vec v$. This will be shown in the following through a flat space example where the piecewise linear string is exhibited as a limit of smooth strings.

\subsection{Collisions in flat space}

In (2+1)-dimensional flat space, an explicit string solution $X(z,\bar z)$ that illustrates this behavior is given by\footnote{See section 2 of \cite{Jevicki:2009uz} for related solutions.}
\bea
  \nonumber
&  f(z) = a_1 \tanh \epsilon (z-z_0) & \\
  \nonumber
&  g(\bar z) = a_2 \tanh \epsilon (\bar z-{\bar z}_0) &
\eea
\bea
  \nonumber
&  \p \vec X(z,\bar z) = \vec e_1 + \half f(z)^2 \vec e_2 - f(z) \vec e_3 & \\
  \nonumber
&  \bar \p \vec X(z,\bar z) = \vec e_2 + \half g(z)^2 \vec e_1 - g(z) \vec e_3 &
\eea

\begin{figure}[h]
\begin{center}
\includegraphics[scale=0.4]{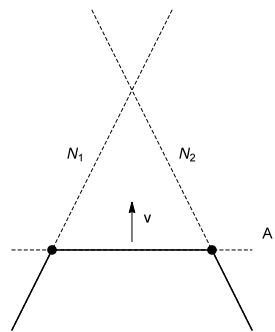}
\caption{\label{fig:expl} Collision of cusps on the string (thick line). In an appropriate frame the lines $N_1$ and $N_2$ are static, whereas $A$ moves with a perpendicular velocity $\vec v$. The collision changes $\vec v \to -\vec v$.
}
\end{center}
\end{figure}

where
\be
  \nonumber
  \vec e_1 = {1 \over \sqrt{2}} (1,1,0);
  \quad   \vec e_2 = {1 \over \sqrt{2}} (1,-1,0);
\quad    \vec e_3 = {1 \over \sqrt{2}} (0,0,1) .
\ee
The spacetime signature is $(-1, 1, 1)$. The $\tanh$ functions can be replaced by any other smooth step functions.

The equations are easily integrated and smooth string solutions are obtained for $\epsilon > 0$ values. The $\epsilon \to 0$ limit is a piecewise linear string: two cusps collide on the worldsheet at $(z,\bar z)= (z_0, \bar z_0)$. The strengths of the cusps are given by $a_1$ and $a_2$.
Additional left- or right-moving cusps can be created by adding extra $\tanh$ functions to $f(z)$ and $g(\bar z)$, respectively. For instance, the solution will have two left-moving cusps if we change
\be
  \nonumber
  f(z) \, \to \,  a_1 \tanh \epsilon (z-z_0) + a_3 \tanh \epsilon (z-z_0')
\ee

A flat space configuration that reproduces the string dynamics in FIG. \ref{fig:expl} can be obtained by setting $z_0 = \bar z_0 = 0$ and $a_1 = a_2 = a$. The string consists of two static halflines and an interval in between that moves with a constant velocity 
\be
\nonumber
\vec v = -{2 \sqrt{2} a \over 2+ a^2} \, \vec e_3.
\ee
The pieces are connected in a smooth way and the smoothing is parametrized by $\epsilon$. After the collision of the two cusps, the velocity of the middle piece changes direction: $\vec v  \to  -\vec v$.

\begin{figure}[h]
\begin{center}
\includegraphics[scale=0.3]{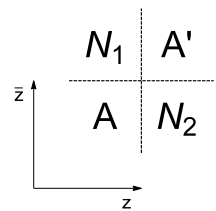}
\caption{\label{fig:latticeexpl} Patches on the worldsheet. The dashed lines in the middle are lightlike worldlines of two colliding cusps. The normal vectors are labeled $\vec A$, $\vec A'$, $\vec N_1$, and $\vec N_2$. The collision formula computes any one of these vectors from the other three. 
}
\end{center}
\end{figure}

\subsection{Collisions in AdS$_3$}

In AdS$_3$, the situation on the worldsheet is shown in FIG. \ref{fig:latticeexpl}. The two dashed lines are the worldlines of the cusps. Before the collision, the string consists of three pieces that are characterized by three normal vectors: $\vec N_1$, $\vec A$ and $\vec N_2$. Note that
\be
  \nonumber
  \vec A \cdot \vec N_1 = \vec A \cdot \vec N_2 = 1 .
\ee
This is required so that the string does not contain badly behaved pieces (see FIG. \ref{fig:lightspeed}). After the collision, $\vec A$ changes to $\vec A'$ given by the {\it collision formula}
\be
  \label{eq:reflection}
  \vec A' = -\vec A +   4 {\vec N_1 + \vec N_2 \over (\vec N_1 + \vec N_2)^2 }
\ee
The formula can be justified by the following facts. It preserves the scalar product:
$  \vec A' \cdot \vec N_1 = \vec A' \cdot \vec N_2 = 1$.
If specialized to the case in FIG. \ref{fig:expl}, the formula reproduces the change in velocity $v \to -v$ for the line $A$. Furthermore, the formula is invariant under $\vec N_1 \leftrightarrow \vec N_2$ and covariant under $SO(2,2)$. Thus, it gives the correct $\vec A'$  in any generic frame. In fact, it computes any one of the ($\vec A$, $\vec A'$, $\vec N_1$, $\vec N_2$) vectors from the other three by an appropriate permutation of the labels. For instance,
\be
  \nonumber
  \vec N_1 = -\vec N_2 +   4 {\vec A + \vec A' \over (\vec A + \vec A')^2 }
\ee
The collision formula can be cast in a Picard-Lefschetz form
\be
  \nonumber
   \vec A' = -\vec A + (\vec A \cdot \vec N ) \vec N \qquad \textrm{with}  \quad
   \vec N = \sqrt{2}{\vec N_1 + \vec N_2 \over |\vec N_1 + \vec N_2|}
\ee
In numerical computations with many cusps, this version introduces exponentially growing numerical errors to the equation $\vec A \cdot \vec N_i = 1$. For such calculations, (\ref{eq:reflection}) is preferable (or a projection has to be performed).

\begin{figure}[h]
\begin{center}
\includegraphics[scale=0.4]{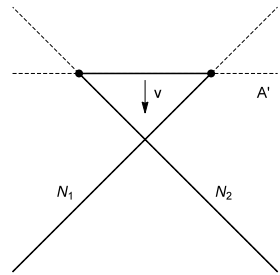}
\caption{\label{fig:explr} The string folds if  $\vec A$ and $\vec A'$ in FIG. \ref{fig:latticeexpl} are exchanged.
}
\end{center}
\end{figure}

Note that one can exchange  $\vec A$ and $\vec A'$ and the constraints on the scalar products of neighbors in FIG. \ref{fig:latticeexpl} are still satisfied. This transformation produces a string embedding that looks like the one in FIG. \ref{fig:explr}: the string is now longer and folded. After the collision, the cusps move away from the open ends of $N_1$ and $N_2$. Thus, swapping the normal vectors reduces the number of future collisions that happen on the \poincare patch.

Finally, for infinitely many weak and tightly spaced cusps, the formula reduces to a differential equation for the normal vector $\vec N(z,\bar z)$
\be
  \nonumber
  \p \bar\p \vec N - (\p \vec N \cdot \bar\p \vec N ) \vec N = 0 .
\ee
This is precisely the same equation as (\ref{eq:eoms}) for $\vec Y$. This is no coincidence: the similarity follows from an internal $SO(2,2)$ symmetry that acts on the $\vec Y, \, e^{-\alpha}\bar\p\vec Y, \, e^{-\alpha}\p\vec Y$, and $\vec N$ variables  \cite{Alday:2009yn}.

\section{Examples}

In this section, a few string solutions are presented. The solutions are based on the circular string in FIG. \ref{fig:basicstring}. Cusps are sent in from the boundary by perturbing both endpoints.

The figures have been generated by a Mathematica \cite{ram2014} code that is available to the reader.

\begin{figure}[h]
\begin{center}
\includegraphics[scale=0.42]{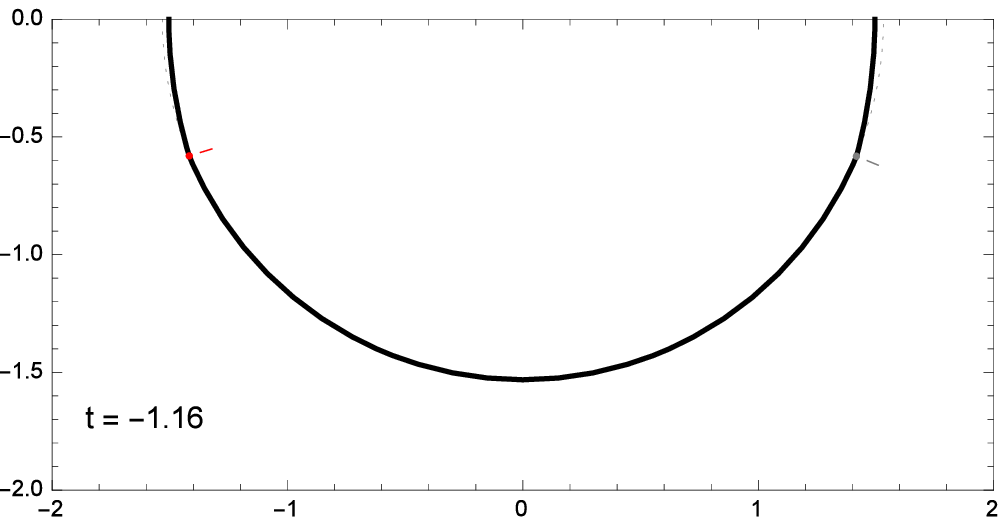} 
\includegraphics[scale=0.42]{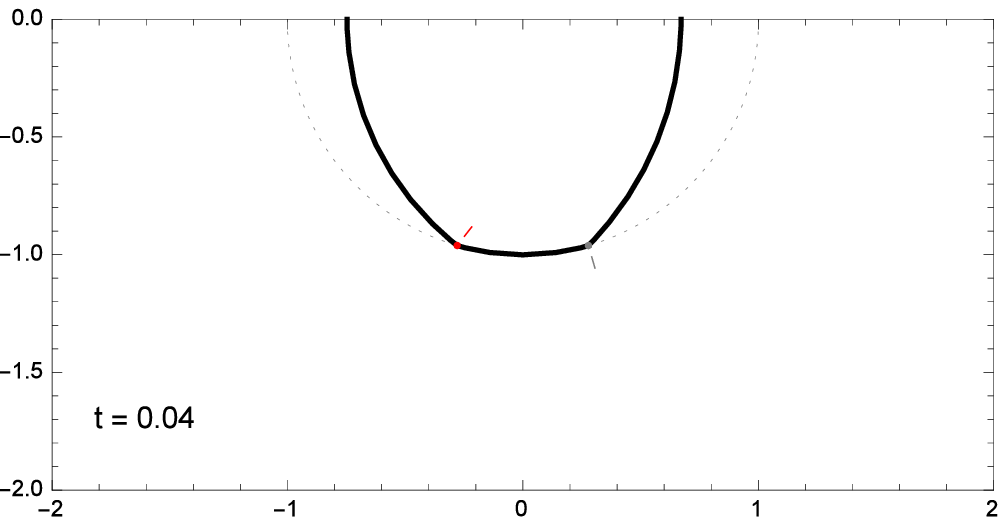} 
\includegraphics[scale=0.42]{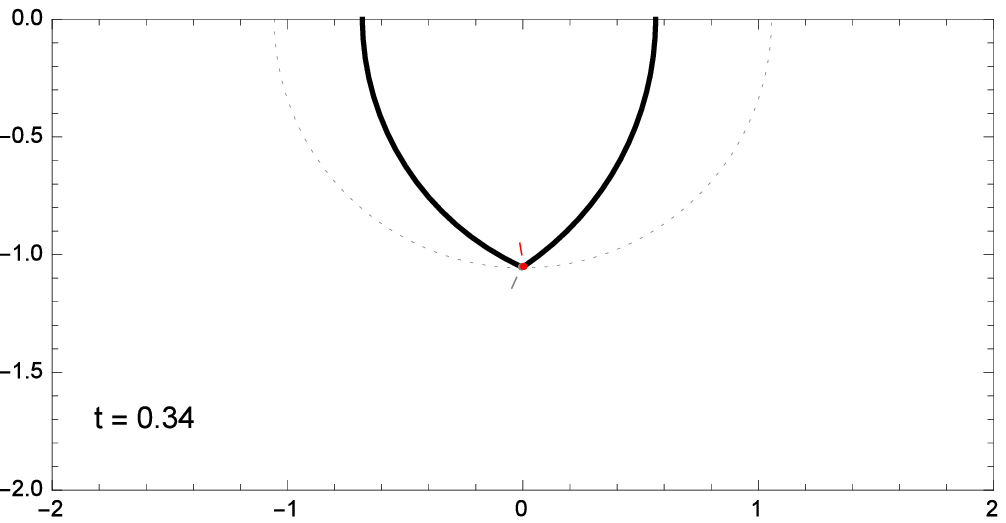} 
\includegraphics[scale=0.42]{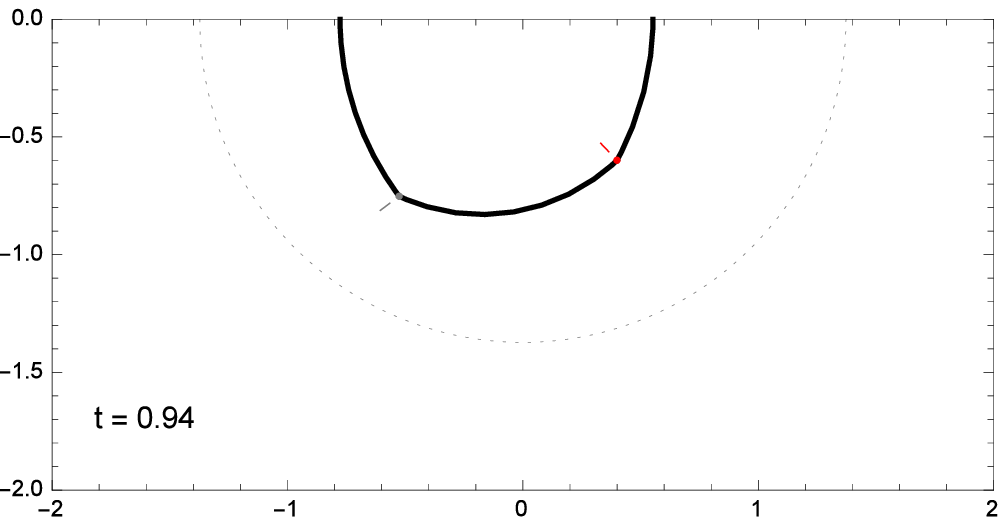}
\caption{\label{fig:cuspcoll} Two cusps (red and gray ticks) colliding on the string (thick line). The figures show the $x-z$ halfplane at various times. The string ends on the boundary of AdS on the top of the figures.
}
\end{center}
\end{figure}

\begin{figure}[h]
\begin{center}
\includegraphics[scale=0.42]{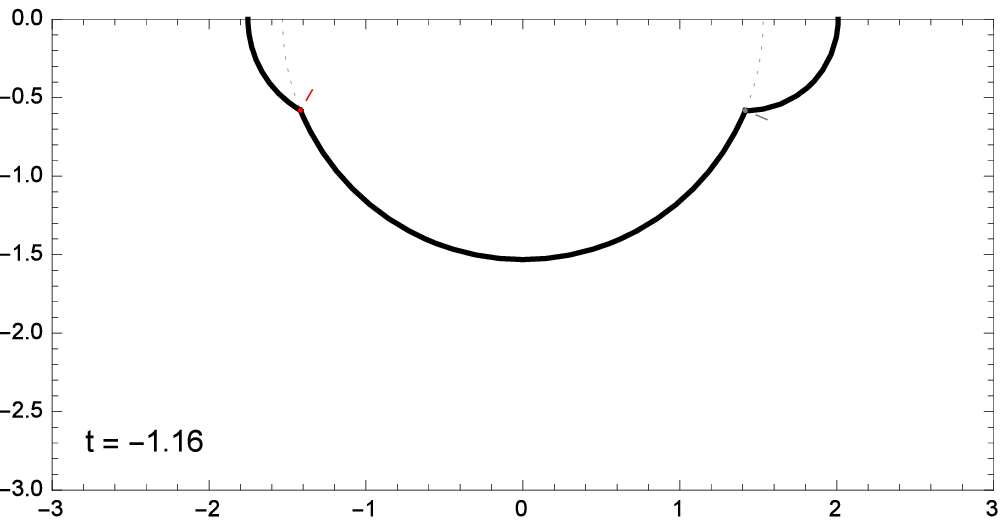} 
\includegraphics[scale=0.42]{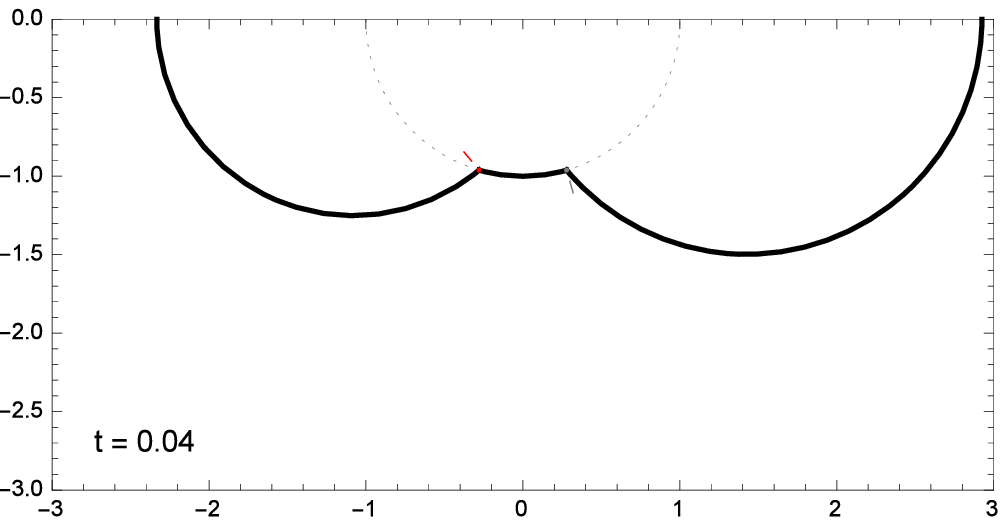} 
\includegraphics[scale=0.42]{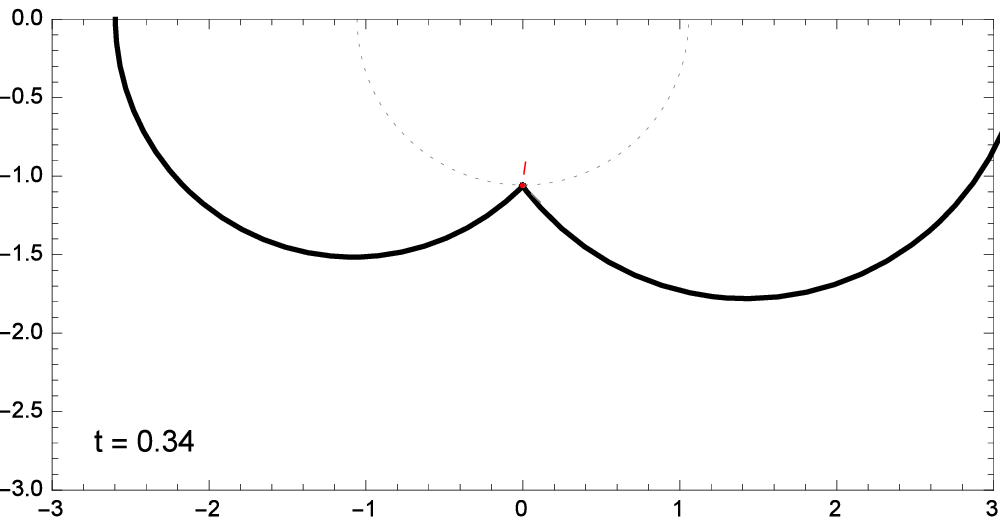} 
\includegraphics[scale=0.42]{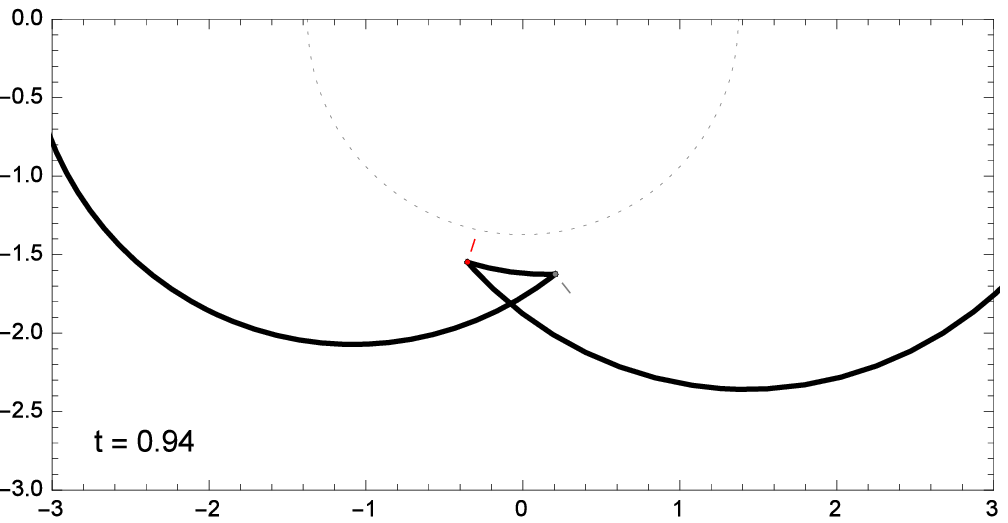}
\caption{\label{fig:cuspcoll2} Two cusps colliding with inverted momenta. The string folds after the collision.
}
\end{center}
\end{figure}

\begin{figure}[h]
\begin{center}
\includegraphics[scale=0.42]{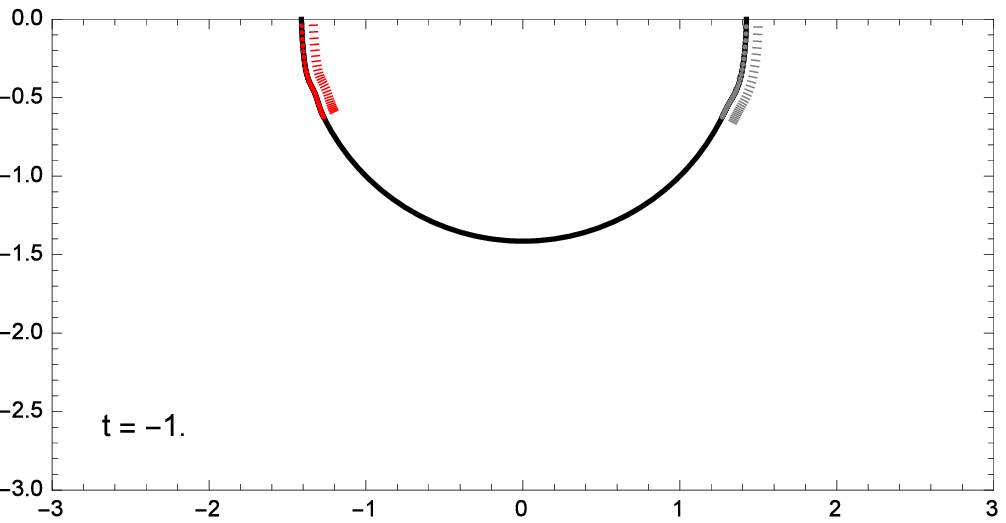} 
\includegraphics[scale=0.42]{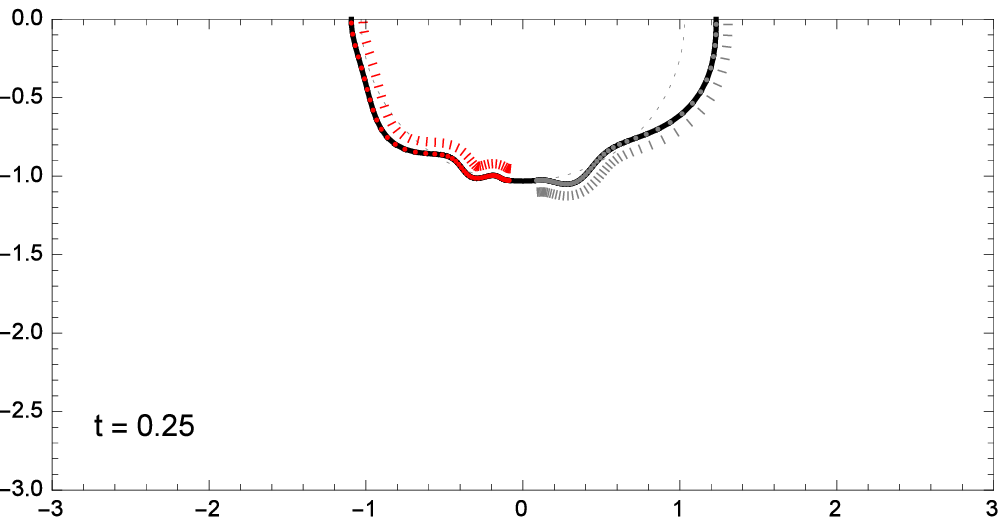} 
\includegraphics[scale=0.42]{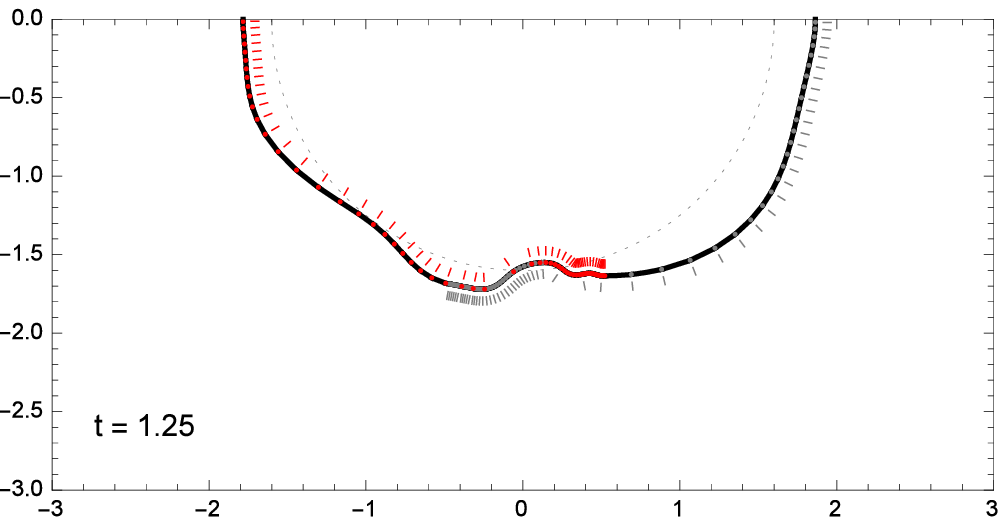} 
\includegraphics[scale=0.42]{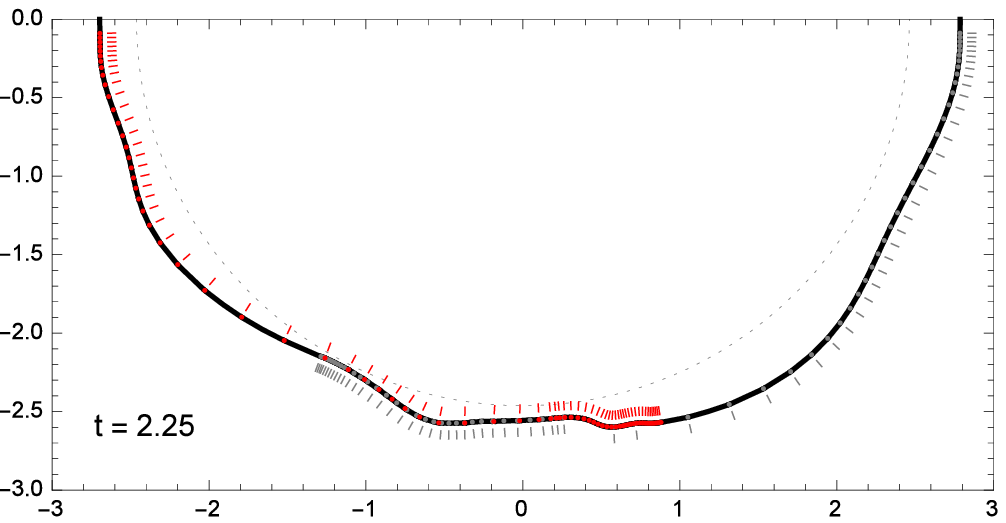}
\caption{\label{fig:manycoll} Collision of 70 left-moving and 70 right-moving cusps. The string has a smooth appearance.
}
\end{center}
\end{figure}

The first example is shown in FIG. \ref{fig:cuspcoll}. There is one left-moving and one right-moving cusp on the worldsheet. They are indicated by red and gray ticks. Dashed circle indicates the original patch. Without the cusps, the string would lie on this circle. The third figure shows the moment of collision. The cusps move through each other. The patch between the two cusps is reflected using the collision formula.

Another example is shown in FIG. \ref{fig:cuspcoll2}. In this case, the cusps have opposite momentum and their presence makes the string longer. After their collision, the string folds and the two cusp angles become large\footnote{Folding is also observed for smooth strings. Cusps are created even if the string was initially smooth.}.

The third example is shown in FIG. \ref{fig:manycoll}. A smooth string is approximated by letting 70 weak cusps enter the string on both sides. The cusp locations are shown in red and gray.

\begin{figure}[h]
\begin{center}
\includegraphics[scale=0.3]{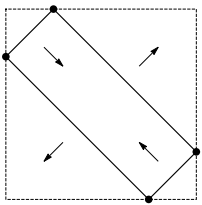}
\caption{\label{fig:closed} A closed string example with four cusps. Arrows show the motion of the different string pieces. The cusps move on the sides of the dashed square with the speed of light.
}
\end{center}
\end{figure}

So far, the examples have been open strings that end on the boundary of AdS. However, closed strings can also be built in a similar fashion. A flat space example is presented in FIG. \ref{fig:closed}. The string (thick line) has two left-moving and two right-moving cusps that move on the dashed square.  After the collision, the cusps start moving in the opposite direction. At any given time, the shape of the string  is a rectangle and the string oscillates.

\section{Discussion}

The string embeddings constructed in this paper can be used to approximate any smooth string motion on the \poincare patch. However, they are also interesting by themselves, because they are exact solutions to the classical equations of motion. Consequently, this type of discretization does not introduce any numerical errors that otherwise might accumulate over time and would lead to various numerical instabilities.

Motion of the string endpoints on the boundary has been specified through the third time derivative of their positions. Cusps on the string were created by adding delta functions to $x'''(t)$. The cusps can be smoothed by resolving the delta functions.

The results can be generalized in various ways. In higher dimensional backgrounds, the collision of cusps can be reduced to the (3+1)-dimensional case. In flat space, the collision event is a deformation of FIG. \ref{fig:expl} where $N_1$ and $N_2$ do not lie in the same plane. It would also be interesting to study brane dynamics based on the techniques presented in this paper.

Another generalization can be the inclusion of an emblackening factor in the background geometry. Integrability of the theory will presumably be lost, but approximate solutions similar to the ones in the present paper may still be of use.  An idea is to exhibit the background geometry as a sum of thin  $AdS_3$ slices with a given $\Delta z$ thickness. Then, as the tiny string patches travel in the $z$ direction, they need to interact with the $AdS_3$ domain walls in some way. One can hope to satisfy the equations of motion in the $\Delta z \to 0$ limit.

The implications of the results for the holographic \cite{Jensen:2013ora} ER=EPR correspondence \cite{Maldacena:2013xja} will be discussed elsewhere \cite{vegh1}.

\vspace{0.2in}   \centerline{\bf{Acknowledgments}} \vspace{0.2in}
The author would like to thank Douglas Stanford for discussions and helpful comments on the manuscript, and the University of Leiden for hospitality.

\clearpage

\appendix
\section*{Appendix}

In this appendix, we discuss some details of the attached Mathematica code.
The code generates plots of the string on the \poincare patch by sending in \texttt{SOLNUM} cusps from both the left and right  endpoints.

First it computes the normal vectors  and stores them in the \texttt{SOLNUM}$\times$\texttt{SOLNUM} matrix \texttt{vectable}. In order to do this, it starts with the string corresponding to the normal vector $\vec N_0 = (0,0,0,1)$. At increasing \poincare time it adds cusps to the left side of the strings and then to the right side of the string and computes the corresponding $\vec N$ vectors. They are stored in the first row and first column of \texttt{vectable}. The ``strengths'' of the cusps are taken from the predefined lists \texttt{lambda1} and
\texttt{lambda2}.
Other elements of \texttt{vectable} are computed by means of the collision formula (\ref{eq:reflection}).

Collision times  are computed and stored in the \texttt{timetable} matrix. Sometimes a collision is calculated to happen earlier than previous collisions. This means that the collision event actually takes place on the next \poincare patch and therefore must be discarded. In this case, the corresponding time in \texttt{timetable} is set to \texttt{CUTOFF} (a large number).

\begin{figure}[h]
\begin{center}
\includegraphics[scale=0.45]{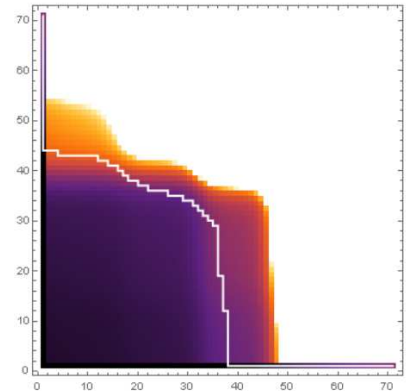}
\caption{\label{fig:lattice} The \texttt{timetable} matrix for the string that is shown in FIG. \ref{fig:manycoll}. The rows and the columns label left- and right-moving cusps on the string, respectively. Colors indicate the time of collision (brighter colors correspond to later times and white means that the collision never takes place on the  \poincare patch). The chain of patches corresponding to the 4th plot in FIG. \ref{fig:manycoll} is shown as a white line. It extends from the top-left corner to the bottom-right corner which means that all 140 cusps have already entered the string.
}
\end{center}
\end{figure}

Finally, \texttt{TIMESTEPS} plots are generated for  \poincare times between \texttt{MINTIME} and \texttt{MAXTIME}. For a fixed  time, the code computes a path in \texttt{vectable} that goes through the relevant string patches. This list of points is stored in \texttt{path}. A sample path is shown in FIG. \ref{fig:lattice} in white. Along the path, the code computes the arcs of strings corresponding to the normal vector of each patch. They are drawn and stored in \texttt{plots}.

\bibliography{broken}

\end{document}